\documentclass[3p,times,procedia]{elsarticle}
\usepackage{ecrc}
\usepackage[bookmarks=false]{hyperref}
    \hypersetup{colorlinks,
      linkcolor=blue,
      citecolor=blue,
      urlcolor=blue}
\volume{00}

\firstpage{1}

\journalname{Procedia Computer Science}

\runauth{Edthofer, Huber et al.}

\jid{procs}

\usepackage{amssymb}
\usepackage{amsmath}
\usepackage[figuresright]{rotating}
\newcommand{\dd}{\mathrm d}   % Differential-d

\begin{document}
\begin{frontmatter}



\dochead{12th. Federation of European Simulation Societies Conference,
(Eurosim 2026)
}%%%
%% e.g. \dochead{17th International Conference on Dynamical Processes in Excited States of Solids}

\title{Introducing entropy measures to PK/PD models in propofol anesthesia as a replacement of BIS}


\author[a]{Alexander Edthofer \corref{cor1}} %\hspace{10 mm}
\author[a]{Christina Huber} 
\author[a,b]{Fabrizio Renzaglia}
\author[a]{Andreas Körner}

\address[a]{Institute of Analysis and Scientific Computing, TU Wien, Wiedner Hauptstraße 8-10, 1040 Vienna, Austria}
\address[b]{Università di Padova, Via VIII Febbraio 2, 35122 Padova, Italy}

\begin{abstract}
Depth of anesthesia is a complex but important vital state to analyze during a surgery or other procedure. One parameter to estimate this state is the bispectral index (BIS), a value ranging from 0 to 100 with a target of 40 to 60 for a stable state during surgery, which is based on the electroencephalogram (EEG). Despite its widespread clinical use, the BIS remains the subject of ongoing discussion as the exact algorithm underlying the BIS is not publicly disclosed, motivating the search for alternative EEG-based indices.
In this publication, two entropy-based EEG measures, Permutation Entropy (PeEn) and Entropy of Difference (EoD) are investigated as potential alternatives to replace the BIS for anesthesia monitoring. Both measures quantify the complexity and irregularity of EEG signals and have previously been proposed as indicators of changes in consciousness and anesthetic depth.
Their performance is evaluated by comparing with the simulated BIS values generated using the pharmacokinetic/pharmacodynamic (PK/PD) models, proposed by Marsh, Schnider and Eleveld. For each model, the root mean squared error (RMSE) between the simulated BIS and the recorded BIS, PeEn, or EoD is calculated, respectively.
Statistical analysis using the Wilcoxon rank-sum test reveal no significant differences between the median RMSE values of the simulated BIS, PeEn, and EoD across the investigated PK/PD models. These results suggest that PeEn and EoD provide performance comparable to that of the simulated BIS, indicating that they may represent promising EEG-based indicators for monitoring depth of anesthesia.
\end{abstract}

\cortext[cor1]{Corresponding author. alexander.edthofer@tuwien.ac.at}

%\correspondingauthor[*]{Corresponding author. Tel.: +0-000-000-0000 ; fax: +0-000-000-0000.}
%\email{alexander.edthofer@tuwien.ac.at}

\end{frontmatter}

% \linenumbers
\vspace*{-6pt}

%\enlargethispage{-7mm}
\section{Introduction}
\label{main}

Maintaining an adequate level of anesthesia during surgery is important to avoid intraoperative awareness on the one hand and to reduce the risk of postoperative delirium on the other. In order to enhance both the efficiency and safety of drug administration, semi-automatic infusion systems, so called infusion pumps have been introduced as part of the target-controlled infusion (TCI) approach described in \cite{al2016principles}. Based on pharmacokinetic/pharmacodynamic (PK/PD) models, these systems aim to estimate the optimal dosage required to achieve a desired target concentration in the blood and help to monitor the depth of an anesthesia.

For the assessment of the vital state of patient different biophysiological signals, such as heart rate via electrocardiogram (ECG), respiratory rate, and electromyogram (EMG) are used. Determining the depth of anesthesia is very difficult. Since 1994, the bispectral index (BIS) (Medtronic, Dublin, Ireland) has been used, which incorporates multiple electroencephalogram (EEG) features for monitoring anesthesia. However, the exact calculation method is unclear. Another disadvantage is that it only works well with middle-aged people, not with older people, especially frail ones, who make up the largest proportion of people undergoing surgery. Therefore, the BIS is topic of much discussion among anesthesiologists. In this publication, we compare anesthesia depth monitoring using PK/PD models with BIS monitoring using entropy-based EEG indices, i.e. permutation entropy (PeEn) and entropy of difference (EoD). As we have previously demonstrated, these indices are effective in estimating different stages of consciousness \cite{edthofer_entropy_2024, edthofer2023permutation} and the BIS correlates with the PeEn \cite{olofsen2008permutation}.

First, in section \ref{sec:data} we will introduce the analyzed dataset. Second, section \ref{sec:PK/PD} presents general PK/PD models and then three specific models developed between 1991 and 2018 that allowed for more and more patient-individualized treatment. Third, in section \ref{sec:entropy}, we will summarize the most important aspects of analyzing an EEG as a timeseries using the PeEn and the EoD. Then, in section \ref{sec:results}, we present our results, followed by a discussion in section \ref{sec:discussion}.

\section{Analyzed dataset} \label{sec:data}
The dataset used for this analysis was obtained from an observational study conducted by Federico Linassi \cite{linassi2023schnider}. The original cohort comprised 78 female patients, including adult patients with age $<65$ years and elderly patients with $\geq 65$ years. Drug administration was guided using either the Eleveld or the Schnider PK/PD model.
The depth of anesthesia was monitored using the BIS. Target values were $40$ and $60$ during maintenance for the patients which underwent breast surgery under propofol–remifentanil total intravenous anesthesia with TCI. Some patients underwent a surgical plethysmographic index for antinociception monitoring without neuromuscular blockade, which had BIS target values of 20 to 50. In addition, EEG time-series recorded from four frontal electrodes were available for analysis.
As part of the EEG preprocessing workflow and the computation of entropy-based measures, patients were screened for the availability and quality of both BIS and EEG recordings. Patients with incomplete BIS data or those who experienced adverse events during anesthesia were excluded, reducing the sample size from $78$ to $59$ individuals. A further three patients were excluded because EEG recordings were unavailable, yielding a study population of $56$ patients.

\section{PK/PD models} \label{sec:PK/PD}
PK/PD models are compartment models that couple linear pharmacokinetics with a with a static nonlinearity to link dose, concentration and drug effect.
The pharmacokinetic part, which describes the changes in drug concentration over time, is represented by a linear multi-compartment model. The PK model is extended by a hypothetical compartment, referred to as the effect-site compartment, which as part of the PD model, accounts for the delay between the plasma concentration of the drug and its corresponding effect. The nonlinear aspect is incorporated through a Hill function, which defines the relationship between the effect site concentration and the clinical effect, measured as the BIS level.

\begin{figure}[ht]\vspace*{4pt}
\centerline{\includegraphics[width = 0.62\textwidth]{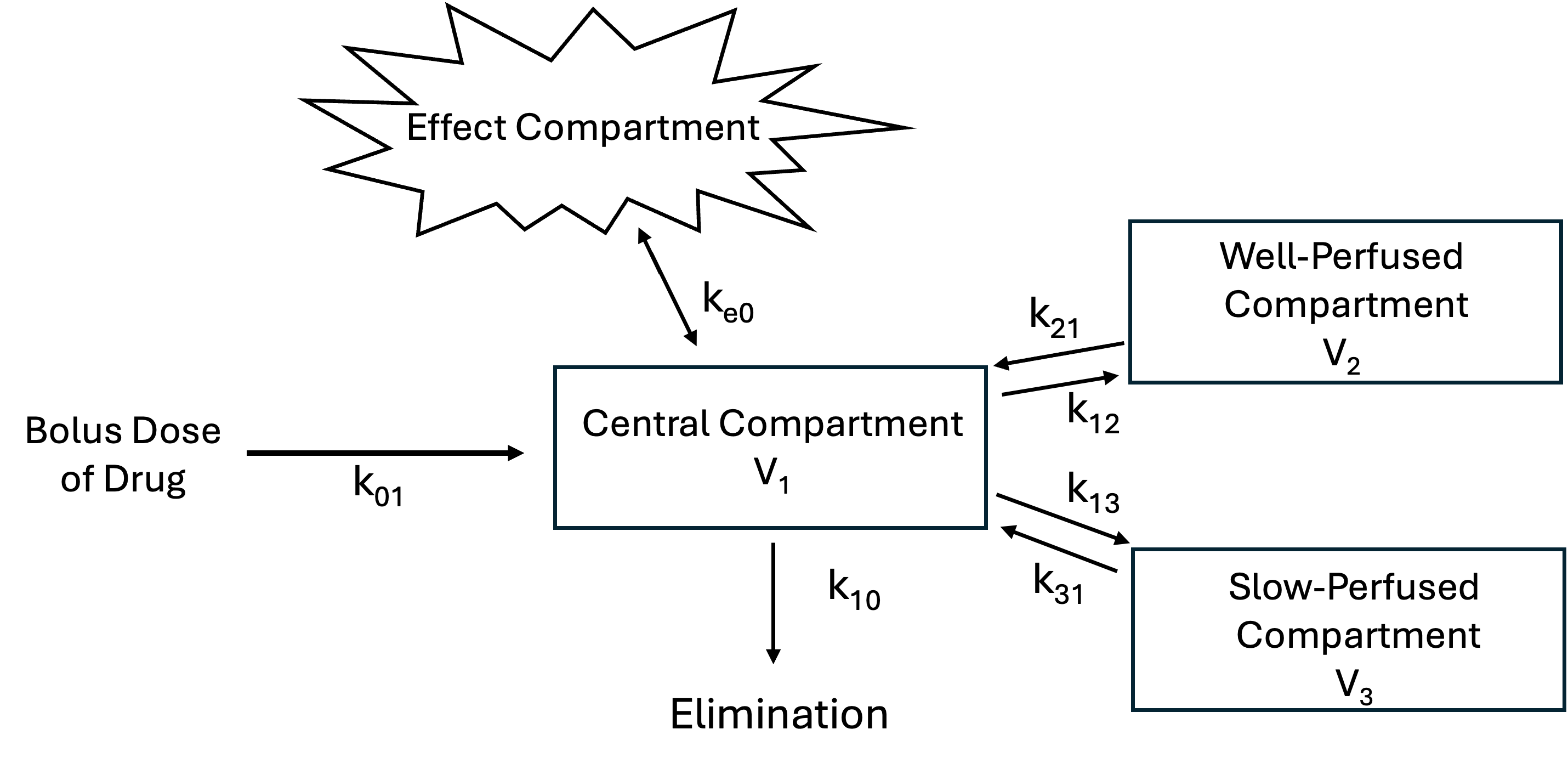}}
\caption{Schematic representation of a PK/PD model consisting of a 3-compartment PK-model extended by the hypothetical effect-side compartment}
\label{fig:pharmacody3comp}
\end{figure}
In the following, we will focus on the PK/PD models proposed by Marsh \cite{marsh1991pharmacokinetic}, Schnider \cite{schnider1998influence} and Eleveld \cite{Eleveld2018}, which are currently implemented in TCI infusion pumps. These models all utilize a three compartment PK/PD model, as illustrated in Figure \ref{fig:pharmacody3comp}, whereby the approaches differ in the underlying assumptions used to calculate the model parameters.
Usually, the first compartment of the PK model, referred to as the central compartment, corresponds to the blood plasma, while the second and third compartment represent peripheral compartments for well-perfused and slow-perfused tissue respectively.
The three compartments with volumes $V_1$, $V_2$ and $V_3$ respectively, are interconnected through mass flux exchange, which is described by the drug transfer frequency $k_{ij}$, given by $$k_{ij} = \frac{CL_j}{V_i} \hspace{3mm} \text{for} \hspace{3mm} i \neq j,$$ where $CL_j$ represents the intercompartmental clearance of compartment $j$. These clearances depend on patient-specific variables such as age, weight or gender, helping to account the influence of these quantities on the rate constants.
This leads to the following system of ordinary differential equations, where $C_i$ represents the concentration of the drug in compartment $i = 1,2,3$ over time.
\begin{align*}
\frac{\dd C_1}{\dd t} &= \frac{u(t)}{V_1} - (k_{10} + k_{12} + k_{13}) C_1(t) + k_{21} C_2(t) \frac{V_2}{V_1} + k_{31} C_3(t) \frac{V_3}{V_1}\nonumber \\ 
\frac{\dd C_2}{\dd t} &= k_{12}  C_1(t) \frac{V_1}{V_2} - k_{21}  C_2(t)\\ 
\frac{\dd C_3}{\dd t} &= k_{13} C_1(t) \frac{V_1}{V_3} - k_{31} C_3(t) \nonumber
\end{align*}
with initial conditions $C_1(0) = \text{bolus}$, $C_2(0) = C_3(0) = 0$ and input function $u(t)$, which represents both the initial bolus dose and the subsequent infusion rate of the drug..
The relation between the concentration in the central compartment and the effect-site concentration $C_e$ is given by
\begin{align*}
\frac{\dd C_e}{\dd t} = k_{e0} C_1(t) - k_{e0} C_e(t).
\end{align*}

\subsection{Marsh model}
Marsh's PK/PD model was one of the first and simplest models of its kind, described in \cite{marsh1991pharmacokinetic}. Unlike the other two models, Marsh incorporates weight as the only co-variate, applying a linear scaling to the volumes of distribution
$$V_i = \theta_i \cdot \text{weight}.$$
The corresponding parameter values and rate constants are given in the corresponding publication \cite{marsh1991pharmacokinetic}. 
The original version of the Marsh model do not account for the $k_{e0}$ rate, nor does it provide any connection for calculating the BIS. 

For our computation we use the same values and formula as Martín-Mateos et al. \cite{martin2013modelling} developed for the Schnider model. This is described in the next section.

\subsection{Schnider model}
Unlike the Marsh model, the PK/PD model proposed by Schnider  \cite{schnider1998influence} already takes several co-variates into account. While the volumes $V_1$ and $V_3$ as well as the intercompartmental clearance $CL_3$, and thus also $k_{13}$ and $k_{31}$, are fixed, $V_2$ and $CL_2$ depend on the age of the patient. The clearance from the central compartment $CL_1$ is influenced by weight, height, and the lean body mass index, which is calculated based on gender as follows.
\begin{align}
\text{lmb} &= \begin{cases} 
1.1 \cdot \text{weight} - 128 \cdot \frac{\text{weight}^2}{\text{height}^2}, & \text{for male.} \nonumber \\
1.07 \cdot \text{weight} - 148 \cdot \frac{\text{weight}^2}{\text{height}^2}, & \text{for female.}
\end{cases}  \nonumber
\end{align}
The final parameters of the Schnider model are given in the corresponding publication \cite{schnider1998influence}.

In order to characterize the relationship between Ce and the BIS in the Schnider PK/PD model, Martín-Mateos et al. \cite{martin2013modelling} describes a sigmoidal $E_{max}$ model, based on a general Hill function, which is commonly used to capture continuous-valued nonlinear PD effects. The BIS is given by
\begin{align}
\label{martin_mateosz_bis}
BIS = BIS_{\text{baseline}} - (BIS_{\text{baseline}} - BIS_{min}) \frac{C_e^\gamma}{C_{e50}^\gamma+C_e^\gamma},
\end{align}
where $BIS_{\text{baseline}}$ describes the baseline BIS in the absence of drug effect i.e the awake state, $BIS_{min}$ stands for the minimum BIS value corresponding to the maximal drug effect, $C_{e50}$ is the effect‑site concentration producing $50\%$ of the maximal effect and the hill coefficient $\gamma$ governs the steepness of the concentration–effect relationship. Martín-Mateos denotes the baseline BIS by $BIS_{0}$. However, to maintain consistency with the following sections, we adopt the notation $BIS_{\text{baseline}}$ used by Eleveld \cite{Eleveld2018}.

Using clinical data from $42$ non-premedicated patients, Martín‑Mateos et al. established two models, one for the induction and one for the maintenance phase, estimating the Hill parameters $C_{e50}$ and $\gamma$. Values for $BIS_{0}$ and $BIS_{min}$ were fixed at $95.6 \pm 4.6$ and $8.9$, respectively, following Vanluchene et al. \cite{vanluchene2004spectral} The effect-side concentration $Ce$ was obtained directly from the Schnider PK/PD model. Based on summary statistics, i.e. mean, standard deviation and the correlation coefficient, $C_{e50}$ and $\gamma$ for both, the induction and maintenance phase, are estimated. The values obtained for the total cohort are provided in section \ref{para_for_BIS}, gender-specific estimates, for male and female, are listed in \cite{martin2013modelling}.

\subsection{Eleveld model}
The PK/PD model of Eleveld \cite{Eleveld2018} is the newest of the three models, incorporating the co-variates weight, height, gender, and age by employing allometric scaling of both, the volumes and the clearance rates between the compartments. This makes the model applicable to a broad population, including adults, children, older adults, and overweight individuals. Moreover, maturation models, incorporating postmenstrual age (PMA), are utilized to reflect the development of drug-eliminating organs, providing insights into how drug effects vary in newborns and children. The central volume $V_1$ scales with weight using a sigmoid function that considers maximum saturation, accounting that larger bodies do not provide unlimited distribution volume. Some parameters also take into account whether there are already anesthetics or other drugs in the bloodstream and draw differences in drug concentration between arterial and venous blood. Inter-individual variability is modeled by incorporating normally distributed random variables $\eta_i$ into the calculation of model parameters. Most of the parameters require scaling relative to a reference individual, which is defined as a 35 year old male person, with a height of 170 cm and weight of 70 kg. The formulas for obtaining the final parameters can again be found in the corresponding publication \cite{Eleveld2018}. 

Again the BIS is modeled using a sigmoidal $E_{max}$ model, driven by the effect compartment concentration $C_e$. Eleveld introduces the BIS via the following equations
\begin{align}\label{BIS_eleveld}
&BIS = BIS_{\text{baseline}} \cdot \left(1 - \frac{C_e^{\gamma}}{C_{e50}^{\gamma} + C_{e}^{\gamma}}\right),\nonumber\\
&BIS_\text{baseline} = \hat\Theta_3.
\end{align}
The drug effect corresponds to the proportion of the maximum achieved drug effect, ranging from $0$ to $1$, i.e. $0\%$ to $100\%$. As in the BIS model introduced by Martín‑Mateos \cite{martin2013modelling}, $C_{e50}$ is the effect-site concentration at which $50\%$ of the maximal drug effect occurs and the $BIS_\text{baseline}$ indicates the BIS value in the absence of drug effect. The quantity $\gamma$ describes the steepness of the concentration-effect relationship. The parameter $\hat\Theta_3$ denotes an estimated model coefficient, all together they are summarized in the section \ref{para_for_BIS_Eleveld}. A detailed explanation of the PD model derivation can be found in the work by Eleveld \cite{Eleveld2018}.

\section{Entropy-based EEG indices} \label{sec:entropy}

Entropy-based indices have gained increasing popularity for analyzing EEG signals during sleep or anesthesia \cite{jordan2008electroencephalographic, liang2015eeg} and have proven particularly effective in separating different states of consciousness \cite{edthofer2023permutation}. In the following, we will outline two such indices, the Permutation entropy (PeEn) and the Entropy of Difference (EoD). Both are based on Shannon's information entropy \cite{shannon1948mathematic}, with PeEn relying on ordinal sequences, while EoD uses patterns in the sign of the difference between two data points.
For a random variable $X=(x_1,... ,x_n)$ with probability distribution $p(x_i), i=1,\ldots,n$, the entropy according to Shannon is defined as
\begin{align}
\label{shannon_entropy}
H(X) = - K \sum_{i=1}^n p(x_i) \log_2 p(x_i).
\end{align}

\subsection{Permutation Entropy}
In order to derive the PeEn, an embedding dimension $m$ and time-delay $\tau$, which specifies the index shift between consecutive elements in a tuple, are first selected. The EEG time series $(x_t)_{t \in I}, I={1,...,N}$ is then divided into  $$k = N-(m-1)\tau$$ tuples of length $m$, serving as the basis for ordinal pattern encoding. Within each tuple, the largest amplitude is assigned the rank $m$, while the smallest amplitude is assigned the rank $1$. This procedure results in $m!$ possible permutations, denoted by $\pi_i$. 
The PeEn value lies in the range $[0, \log(m!)]$. The minimum value of $0$ is obtained when a single ordinal pattern appears in every tuple. In contrast, the maximum value $\log(m!)$ is reached when all permutation types occur with equal probability, i.e., when the distribution of ordinal patterns is uniform $p(\pi_i) = \frac{1}{m!}$. To achieve the maximum value, the number of tuples must be at least equal to the number of possible permutations, i.e. $k \geq m!$.
In terms of Shannon's entropy formula (\ref{shannon_entropy}) the normalized measure of the PeEn with a range of $[0, 1]$ is given through
\begin{align*}
PeEn =  - \frac{1}{\log_2 m!}\sum_{i=1}^{m!} p(\pi_i)\log_2 p(\pi_i),
\end{align*}
where $p(\pi_i)$ describes the probability of occurrence of permutation $i$.

A more detailed description of the PeEn Algorithm can be found in the work of Jordan et al. \cite{jordan2008electroencephalographic} or Bandt and Pompe. \cite{bandt2002permutation}.

\subsection{Entropy of Difference}
For EoD, described in the publication of Pasquale Nardone \cite{nardone2024entropy}, the entropy values are determined not based on ordinal patterns, but by the differences between consecutive amplitudes. When the EEG time series is divided into tuples of length $m$, the focus is placed on the differences between successive samples, or more specifically, on the sign of these differences. Considering the $m$-tuple $[4,2,1,2]$ the differences between the values are given through $[-2,-1,1]$ leading to the tuple of signs $[-,-,+]$. This leads to $2^{m-1}$ possible combinations $\delta_l$ of $"+"$ and $"-"$ signs.

\begin{figure}[ht]%\vspace*{4pt}
\centerline{\includegraphics[width = 0.62\textwidth]{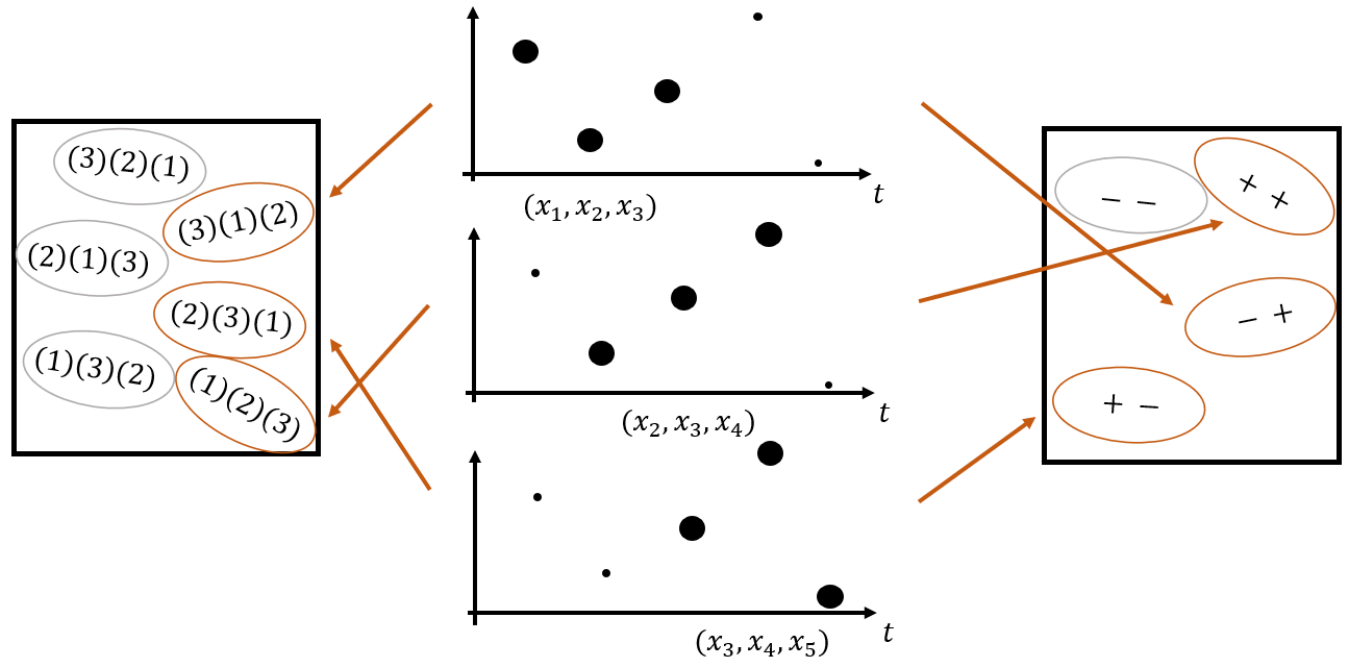}}
\caption{An exemplary time series is encoded once for the PeEn on the left side and once for the EoD on the right side. Figure is taken from \cite{edthofer_entropy_2024}.}
\label{fig:PeEn_EoD}
\end{figure}

The normalized EoD can be expressed in terms of Shannon entropy (\ref{shannon_entropy}) by the formula
\begin{align*}
EoD =  -\frac{1}{m-1} \sum_{l=1}^{2^{m-1}} p(\delta_l)\log_2 p(\delta_l).
\end{align*}
The normalization yields a range of $[0,1]$. Without applying a normalization the range of EoD is given by $[0,{m-1}]$. As with PeEn the minimum value of $0$ is obtained when only a single pattern occurs throughout the entire time series, whereas the maximum value is reached when all patterns occur with equal probability. In contrast to PeEn the number of possible patterns is considerably smaller, which also reduces the amount of tuples required, i.e. $k \geq 2^{m-1}$, for a reliable entropy estimation compared to PeEn.

Figure \ref{fig:PeEn_EoD} illustrates the concepts of PeEn and EoD, showing how EEG time series are decomposed into specific patterns for analysis in the time domain.

\section{Results} \label{sec:results}

\begin{figure}[hb]
\centerline{\includegraphics[width = \textwidth]{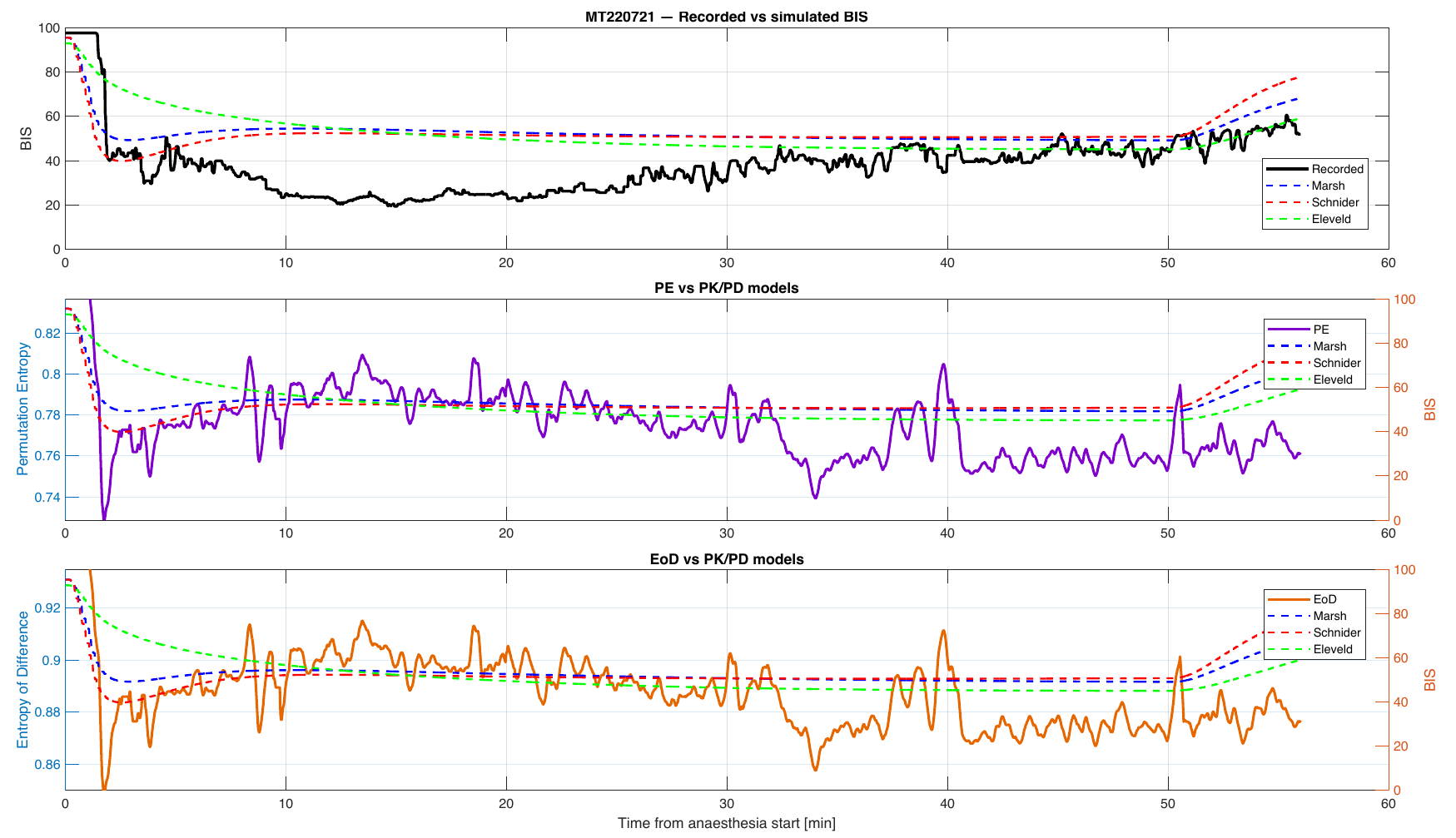}}
\caption{The simulated BIS of the Marsh, Schnider, and Eleveld model are shown in a blue, red and green dashed line respectively in each of the three plots. The $x$-axis shows the time of surgery, starting from the moment the anesthesia started. The top plot shows the comparison to the recorded BIS in black, the middle one the comparison to the PeEn in violet and the bottom one the comparison to the EoD in orange.}
\label{fig:BIS_curves}
\end{figure}

We compare the simulated BIS calculated with formula (\ref{martin_mateosz_bis}) for the Marsh and Schnider models and formula (\ref{BIS_eleveld}) for the Eleveld model with the recorded BIS, the PeEn and EoD respectively. The PeEn and EoD are computed of order $m=3$ of the preprocessed EEG, artifacts are designated as NaN values. An example for the relationship between the simulated BIS-values of one patient and the recorded or calculated measures is shown in Figure \ref{fig:BIS_curves}. The different measures deviate from the simulated BIS at different times. The resemblance of PeEn and EoD is emerging, which we already showed in \cite{edthofer_entropy_2024}. The trajectory of the simulated BIS with Marsh and Schnider is very similar, as both base on the same formula, whereas the one of Eleveld has a different shape.
%\vspace{-8 mm}

For an analysis on the population level we investigate the root mean squared error (RMSE) between the simulated BIS of each model, which was introduced in section \ref{sec:PK/PD}, and the recorded BIS, the calculated PeEn and EoD respectively. As the BIS is always in the range of $[0,100]$, we have to adjust the entropy values accordingly. Therefore, we normalize the respective entropy value in the range $[0,1]$ by 
\[Ent_{new} = \frac{Ent_{old}-\text{min}(Ent_{old})}{\text{max}(Ent_{old})- \text{min}(Ent_{old})},\] 
and then apply a linear transformation. To accomplish this, we set the scaled entropy to be 
\[Ent_{scaled, M\&S} = 69 \cdot Ent_{new} + 26.6, \qquad \text{and} \qquad Ent_{scaled, E} = 76.5 \cdot Ent_{new} + 16.5.\]
Here, $Ent_{scaled, M\&S}$ describes the scaled entropy for Marsh and Schnider, and $Ent_{scaled, E}$ describes the scaled entropy for Eleveld. This is done such that the respective baseline BIS values of 95.6 and 93 for each corresponding model are reached with a minimal entropy value of 26.6 or 16.5.

\begin{figure}[h]\vspace*{4pt}
\centerline{\includegraphics[width = 0.7\textwidth]{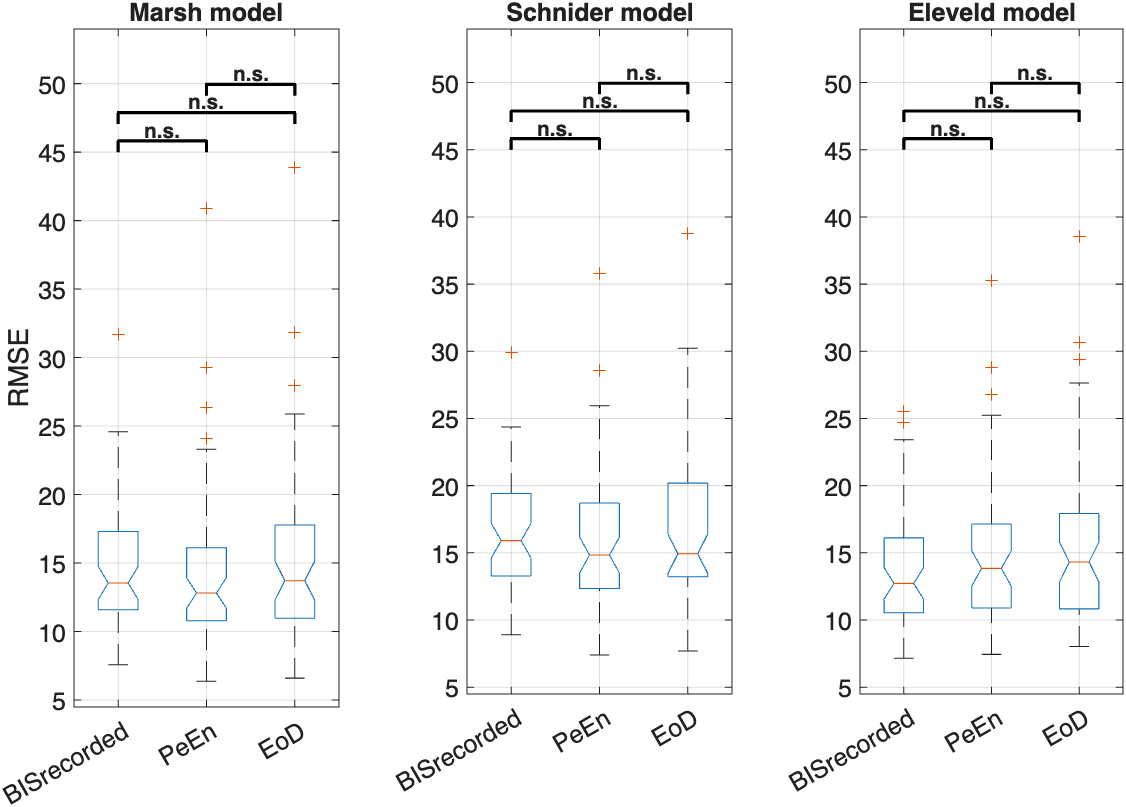}}
\caption{The boxplot diagrams show the root mean squared error (RMSE) between the recorded BIS, calculated PeEn, and EoD with the simulated BIS of the Marsh, Schnider and Eleveld model respectively. The PeEn and EoD are scaled, s.t. the range matches the range of the BIS. The results are not significant, i.e. the medians between groups are approximately the same. This is shown by the "n.s." on top of each plot.}
\label{fig:Boxplot_RMSE}
\end{figure}

We also compute the Wilcoxon rank-sum test to investigate if there is a significant difference between the medians of each measurement for each of the PK/PD models. The corresponding $p$-values are given in Table \ref{tab:p-values}. Because all $p$-values are greater than 0.05, falling within the non-significant range, a Holm-Bonferroni correction was not applied. To visualize the results we use boxplot diagrams, shown in Figure \ref{fig:Boxplot_RMSE}. The results show that there is no significant difference between each measurement, indicating that the median RMSE is the same. This underlines the fact that the PeEn correlates with the BIS \cite{olofsen2008permutation}.

\begin{table}[h] 
\caption{The table shows the $p$-values of the Wilcoxon rank-sum test of the three measurements recorded BIS, PeEn, and EoD, and the three PK/PD models, respectively. As all $p$-values are greater than 0.05, there is no significant difference between the RMSE medians of each measurement.}
\begin{tabular*}{\hsize}{@{\extracolsep{\fill}}llll@{}}
\toprule
\textbf{$p$-values} & Marsh & Schnider & Eleveld \\
\colrule
BISrecorded $\leftrightarrow$ PeEn & 0.2386 & 0.4169 & 0.2727 \\
BISrecorded $\leftrightarrow$ EoD & 0.7074 & 0.8774 & 0.1102 \\
PeEn $\leftrightarrow$ EoD  & 0.4581 & 0.5587 & 0.5316 \\
\botrule
\end{tabular*}
\label{tab:p-values}
\end{table}

\section{Discussion} \label{sec:discussion}

The results above show that both the PeEn and the EoD demonstrate performance comparable to that of the simulated BIS. This could motivate the development of EEG-based indices for monitoring the depth of anesthesia.
Simulating entropy values as a function of the effect-site concentration $C_e$ may provide a promising framework for incorporating patient-specific characteristics, such as age, sex, and the ASA score into depth-of-anesthesia monitoring.
Introducing a patient-specific PeEn baseline, analogous to the baseline calibration commonly used for BIS monitoring, could be another potential improvement. Establishing an individual reference value prior to anesthesia induction could reduce inter-individual variability and allow entropy based indices to more accurately reflect changes in brain activity relative to each patient's physiological state. This would support a transition from reliance on a single population based index toward more individualized EEG monitoring approaches, potentially enabling greater use of the full patient-specific spectral information contained in the EEG.

Since the study cohort consisted exclusively of female patients, the applicability of the results to male patients remains to be validated. However, given that the entropy measures are based on fundamental EEG dynamics, similar performance is expected across both sexes.

\section*{Acknowledgments}

We would like to express our sincere gratitude to Matthias Kreuzer and Duygu Aidin from the Neuromonitoring Lab at the Technical University of Munich, Germany, for their insightful suggestions and guidance that have greatly enriched this work. We also want to thank Federico Linassi from the University-Hospital of Padova, Italy, for his valuable support in providing the data for this contribution.


\appendix

\section{Parameters for the BIS modeling by Martín-Mateos et al.} \label{para_for_BIS}

Table \ref{tab:martin} gives the parameters for the calculation of the BIS according to \cite{martin2013modelling}.
\vspace{-4 mm}
\begin{table}[h] 
\caption{Estimated values for $C_{e50}$ and $\gamma$ during induction and maintenance taken from \cite{martin2013modelling}. Values are mean.} \label{tab:martin}
\begin{tabular*}{\hsize}{@{\extracolsep{\fill}}lll@{}}
\toprule
\textbf{Parameter} & \textbf{Induction} & \textbf{Maintenance}  \\
\colrule
$C_{e50}$ & 3.35 & 2.23\\ 
$\gamma$ & 1.24 & 1.58 \\ 
\botrule
\end{tabular*}
\end{table}

\section{Parametertable for the Eleveld model} \label{para_for_BIS_Eleveld}

Table \ref{tab:eleveld} gives the parameters for the calculation of the BIS according to \cite{Eleveld2018}.
\vspace{-4 mm}
\begin{table}[h] 
\caption{Parameters $\hat\Theta_i$ for BIS calculation of the Eleveld model taken from \cite{Eleveld2018}.} \label{tab:eleveld}
\begin{tabular*}{\hsize}{@{\extracolsep{\fill}}llll@{}}
\toprule
\textbf{Parameter} & \textbf{Interpretation} & \textbf{Value} & \textbf{Units}\\
\colrule
$\hat\Theta_1$ & $C_{e50}$ & 3.08 & $\mu \text{g} \cdot \text{ml}^{-1}$\\
$\hat\Theta_2$ & $k_{e0}$ \text{for arterial samples}  & 0.146 & $\text{min}^{-1}$  \\  
$\hat\Theta_3$ & \text{baseline BIS Value}  & 93.0 &\\ 
$\hat\Theta_4$ & \text{PD sigmoid slope ($C_{e} > C_{e50}$)} & 1.74 & \\ 
$\hat\Theta_5$ & \text{Residual error}  & 8.03 & BIS \\ 
$\hat\Theta_6$ & \text{increase in delay with age} & 0.0517 &\\ 
$\hat\Theta_7$ & \text{decrease in $C_{e50}$ with age} &  -0.00635 &\\ 
$\hat\Theta_8$ & \text{$k_{e0}$ for venous samples} & 1.24 & $\text{min}^{-1}$ \\ 
$\hat\Theta_9$ & \text{PD sigmoid slope ($C_{e} \leq C_{e50}$)} & 1.89 &\\ 
\botrule
\end{tabular*}
\end{table}

%% References

%The citation must be used in following style: \cite{article-minimal} \cite{article-full} \cite{article-crossref} \cite{whole-journal}.

\bibliography{literature}
\bibliographystyle{elsarticle-harv}


\clearpage

\end{document}